\pdfoutput=1
\documentclass[12pt]{article}
\usepackage{graphicx}
\usepackage{subcaption}
\usepackage{amssymb}
\usepackage{amsmath}
\usepackage{amsfonts}
\usepackage{multirow}

\usepackage{xcolor}
\usepackage{url}
\usepackage{ulem}
\usepackage{float}
\usepackage{float}

\usepackage{pifont}

\usepackage{soul}
\usepackage[english]{babel}
\usepackage[T1]{fontenc}
\usepackage{lmodern}
\usepackage[utf8]{inputenc}  
\usepackage{bbm}

\usepackage{hyperref}
\hypersetup{
  colorlinks   = true, 
  urlcolor     = blue, 
  linkcolor    = black, 
  citecolor   = black 
}

\usepackage[numbers,sort&compress]{natbib}
\graphicspath{{plots/}}

\newcommand{\lsim}
{\;\raisebox{-.3em}{$\stackrel{\displaystyle <}{\sim}$}\;}
\newcommand{\gsim}
{\;\raisebox{-.3em}{$\stackrel{\displaystyle >}{\sim}$}\;}

\newcommand\tb{\tan\beta}

\newcommand\ReDiag{\mathop{%
  \raise .5pt\hbox{[}%
  \widetilde{\mathrm{Re}}%
  \raise .5pt\hbox{]}}}
\newcommand\ReOffDiag{\mathop{%
  \raise .5pt\hbox{$\llbracket$}%
  \widetilde{\mathrm{Re}}%
  \raise .5pt\hbox{$\rrbracket$}}}

\newcommand\MZ{M_Z}
\newcommand\Mh{M_h}

\newcommand\MA{M_A}

\newcommand\Sn{\tilde\nu}
\newcommand\Sl{\tilde l}

\newcommand\Slpm{\tilde l^\pm}

\newcommand\Sel[1]{\tilde e_{#1}}

\newcommand\mse[1]{m_{\Sel{#1}}}
\newcommand\msl[1]{m_{\Sl_{#1}}}

\newcommand\Stau[1]{{\tilde\tau_{#1}}}
\newcommand\stau{\tilde \tau}

\newcommand\mL{m_{\tilde l_L}}
\newcommand\mR{m_{\tilde l_R}}

\newcommand\ino[1]{\tilde\chi_{#1}}

\newcommand\chapm[1]{\ino{#1}^\pm}

\newcommand\cha{\chapm}
\newcommand\mcha[1]{m_{\chapm{#1}}}

\newcommand\neu[1]{\ino{#1}^0}
\newcommand\mneu[1]{m_{\neu{#1}}}

\newcommand\refeq[1]{Eq.~(\ref{#1})}

\newcommand\refse[1]{Sect.~\ref{#1}}

\newcommand\citere[1]{Ref.~\cite{#1}}
\newcommand\citeres[1]{Refs.~\cite{#1}}

\newcommand{\CP}{{\cal CP}}
\newcommand{\cp}{{\CP}}

\newcommand{\tev}{\,\, \mathrm{TeV}}
\newcommand{\gev}{\,\, \mathrm{GeV}}

\newcommand\MO{\texttt{MicrOMEGAs}}
\newcommand\CM{\texttt{CheckMATE}}

\newcommand\fb{\ensuremath{\mbox{fb}}}
\newcommand\ab{\ensuremath{\mbox{ab}}}

\newcommand\ifb{\ensuremath{\fb^{-1}}}
\newcommand\iab{\ensuremath{\ab^{-1}}}

\newcommand\msmu[1]{m_{\tilde{\mu}_{#1}}}
\newcommand\mstau[1]{m_{\tilde{\tau}_{#1}}}

\newcommand{\sig}{\sigma}

\def\order#1{\ensuremath{{\cal O}(#1)}}
\def\reffi#1{\mbox{Fig.~\ref{#1}}}

\def\ga{\gamma}
\def\De{\Delta}

\def\gmin2{\ensuremath{(g-2)_\mu}}
\def\amu{\ensuremath{a_\mu}}

\newcommand{\ssi}{\ensuremath{\sig_p^{\rm SI}}}

\definecolor{Orange}{named}{orange}
\definecolor{Purple}{named}{purple}
\definecolor{Lightblue}{cmyk}{0.9,0.1,0.1,0.3}
\definecolor{dgelborange}{cmyk}{0.,0.3,0.5, 0.}
\definecolor{Lila}{rgb}{0.5,0.,1}
\definecolor{Darkgreen}{rgb}{0.,.7,0.2}

\newcommand{\fnalcen}{\htrs{204.0}}    
\newcommand{\fnalunc}{\htrs{5.4}}      
\newcommand{\fnalbnlsig}{\htrs{0.8}}   
\newcommand{\newcen}{\htrs{206.1}}     
\newcommand{\newunc}{\htrs{4.1}}       
\newcommand{\newdiff}{\htrs{25.1}}    
\newcommand{\newdiffunc}{\htrs{5.9}}   
\newcommand{\newdiffsig}{\htrs{4.2}}   
\newcommand{\htrs}[1]{{\color{black} #1}}

\evensidemargin \oddsidemargin
\allowdisplaybreaks
\begin{document}
\thispagestyle{empty}

\def\thefootnote{\fnsymbol{footnote}}

\begin{flushright}
\mbox{}
IFT--UAM/CSIC--21-033\\
\end{flushright}

\vspace{0.5cm}

\begin{center}

{\large\sc 
{\bf The new ``MUON G-2'' Result and Supersymmetry
}}

\vspace{1cm}

{\sc
Manimala Chakraborti$^{1}$%
\footnote{email: mani.chakraborti@gmail.com}%
, Sven Heinemeyer$^{2,3,4}$%
\footnote{email: Sven.Heinemeyer@cern.ch}%
~and Ipsita Saha$^{5}$%
\footnote{email: ipsita.saha@ipmu.jp}%
}

\vspace*{.7cm}

{\sl
$^1${Astrocent, Nicolaus Copernicus Astronomical Center of the Polish Academy of Sciences,
ul. Rektorska 4, 00-614 Warsaw, Poland}

\vspace*{0.1cm}

$^2${Instituto de F\'isica Te\'orica (UAM/CSIC),
Universidad Aut\'onoma de Madrid, \\
Cantoblanco, 28049, Madrid, Spain}

\vspace*{0.1cm}

$^3$Campus of International Excellence UAM+CSIC, 
Cantoblanco, 28049, Madrid, Spain 

\vspace*{0.1cm}

$^4$Instituto de F\'isica de Cantabria (CSIC-UC), 
39005, Santander, Spain
\vspace*{0.1cm}

$^5$Kavli IPMU (WPI), UTIAS, University of Tokyo, Kashiwa, Chiba 277-8583, Japan
}

\end{center}

\vspace*{0.1cm}

\begin{abstract}
\noindent
The electroweak (EW) sector of the Minimal Supersymmetric Standard Model
(MSSM), with the lightest neutralino as Dark Matter (DM) candidate, can
account for a variety of experimental data. This includes the DM content
of the universe, DM direct detection limits, EW SUSY searches at the LHC
and in particular the so far persistent $3-4\,\sig$ discrepancy between the
experimental result for the anomalous magnetic moment of the muon, \gmin2, and
its Standard Model (SM) prediction.
The recently published ``MUON G-2'' result is within
$\fnalbnlsig\,\sig$ in agreement with  the older BNL result on \gmin2.
The combination of the two results was given as
$\amu^{\rm exp} = (11 659 \newcen \pm \newunc) \times 10^{-10}$,
yielding a new deviation from the SM prediction of
$\De\amu = (\newdiff \pm \newdiffunc) \times 10^{-10}$, corresponding to
$\newdiffsig\,\sig$. Using this improved bound we update the results
presented in \citere{CHS1} and set new upper limits on the allowed
parameters space of the EW sector of the MSSM.
We find that with the new \gmin2\ result the upper limits
on the (next-to-) lightest SUSY particle are
in the same ballpark as previously, yielding updated upper limits on these
masses of $\sim \htrs{600} \gev$. 
In this way, a clear target is confirmed
for future (HL-)LHC EW searches, as well as for future high-energy
$e^+e^-$~colliders, such as the ILC or CLIC. 
\end{abstract}


\def\thefootnote{\arabic{footnote}}
\setcounter{page}{0}
\setcounter{footnote}{0}

\newpage


\section{Introduction}
\label{sec:intro}

Searches for Beyond the Standard Model (BSM) particles are performed
directly, such as at the LHC, or indirectly in low-energy experiments
and via astrophysical measurements. 
Among the BSM theories under consideration the Minimal Supersymmetric
Standard Model  
(MSSM)~\cite{Ni1984,Ba1988,HaK85,GuH86} is one of the leading candidates.
Supersymmetry (SUSY) predicts two scalar partners for all SM fermions as well
as fermionic partners to all SM bosons. 
Contrary to the case of the SM, in the MSSM two Higgs doublets are required.
This results in five physical Higgs bosons instead of the single Higgs
boson in the SM.  These are the light and heavy $\cp$-even Higgs bosons, 
$h$ and $H$, the $\cp$-odd Higgs boson, $A$, and the charged Higgs bosons,
$H^\pm$.
The neutral SUSY partners of the (neutral) Higgs and electroweak (EW) gauge
bosons gives rise to the four neutralinos, $\neu{1,2,3,4}$.  The corresponding
charged SUSY partners are the charginos, $\cha{1,2}$.
The SUSY partners of the SM leptons and quarks are the scalar leptons
and quarks (sleptons, squarks), respectively.

In \citere{CHS1} we performed an analysis taking into account all
relevant data for the EW sector of the MSSM, assuming that the lightest
SUSY particle (LSP) is given by the lightest neutralino, $\neu1$, that
makes up the full Dark Matter (DM) content of the
universe~\cite{Go1983,ElHaNaOlSr1984}.%
\footnote{
More recently in \citere{CHS2} we updated the analysis using the DM relic density only as an upper bound.
}%
~The expermental results comprised the direct searches at the
LHC~\cite{ATLAS-SUSY,CMS-SUSY}, the DM relic abundance~\cite{Planck},
the DM direct detection (DD) experiments~\cite{XENON,LUX,PANDAX} and in
particular the (then current) deviation of the anomalous magnetic moment
of the muon~\cite{Keshavarzi:2019abf,Davier:2019can}.

Three different scenarios were analyzed, classified by the mechanism
that brings the LSP relic density into agreement with the measured
values. The scenarios differ by the Next-to-LSP (NLSP), where we
investigated $\cha1$-coannhiliation, $\Slpm$-coannihilation with either
``left-'' or ``right-handed'' sleptons close in mass to the LSP, Case-L
and Case-R, respectively. Using the then current bounds we found
upper limits on the LSP masses.
While for $\cha1$-coannihilation this is $\sim 570 \gev$, for
$\Slpm$-coannihilation Case-L $\sim 540 \gev$ and for Case-R
values up to $\sim 520 \gev$ are allowed. Similarly,
upper limits to masses of the coannihilating SUSY particles are found as,
$\mcha1 \lsim 610 \gev$, 
$\msl{L} \lsim 550 \gev$, 
$\msl{R} \lsim 590 \gev$.
For the latter, in the
$\Slpm$-coannihilation case-R, the upper limit on the
lighter $\stau$ is even lower, $\mstau2 \lsim 530 \gev$.
The then current \gmin2\ constraint also yields limits on the rest of the EW
spectrum, although much loser bounds were found. Previous articles with
similar, but in general less advanced studies can be found in
\citeres{Bharucha:2013epa,Fowlie:2013oua,Han:2013kza,Kowalska:2015zja,Choudhury:2016lku,Datta:2016ypd,Chakraborti:2017vxz,Hagiwara:2017lse,Yanagida:2020jzy,Yin:2016shg,Yanagida:2016kag,Chakraborti:2017dpu,Bagnaschi:2015eha,Datta:2018lup,Cox:2018qyi,Abdughani:2019wai,Endo:2020mqz,Pozzo:2018anw,Athron:2018vxy,Carena:2018nlf}.

Recently the ``MUON G-2'' collaboration published the results (referred
to as ``FNAL'' result) of their Run~1 data,which is within
$\fnalbnlsig\,\sig$ in agreement with  the older BNL result on \gmin2.
We combine the two results, assuming that they are uncorrelated. 
We analyze the impact of the combination of the Run~1 FNAL 
data with the previous BNL result on the allowed
MSSM parameter space. The results will be discussed in the context of
the upcoming searches for EW particles at the HL-LHC. We will also comment
on the discovery prospects for these particles at possible future
$e^+e^-$~colliders, such as the ILC~\cite{ILC-TDR,LCreport} or
CLIC~\cite{CLIC,LCreport}.


\section {The model and the experimental constraints}
\label{sec:model-constraints}

\subsection{The model}
\label{sec:model}

A detailed description of the EW sector of the MSSM can be found
in \citere{CHS1}. Here we just list the relevant input parameters and
masses that are relevant for our analysis. 
Throughout this
paper we assume that all parameters are real, i.e.\ we have no
$\CP$-violation.

The masses and mixings of the charginos and neutralinos are determined
by $U(1)_Y$ and $SU(2)_L$ gaugino masses $M_1$ and $M_2$, the Higgs
mixing parameter $\mu$ and $\tb$, the ratio of the two
vacuum expectation values (vevs) of the two Higgs doublets of MSSM,
$\tb = v_2/v_1$.
The four neutralino masses are given as $\mneu1 < \mneu2 < \mneu3 <\mneu4$.
Similarly the two chargino-masses are denoted as $\mcha1 < \mcha2$.
As argued in \citere{CHS1} it is sufficient for our analysis to focus
on positive values for $M_1$, $M_2$ and $\mu$. 

For the sleptons, as in \citere{CHS1}, we choose common soft
SUSY-breaking parameters for all three generations, $\mL$ and $\mR$. 
We take the trilinear coupling
$A_l$ ($l = e, \mu, \tau$) to be zero for all the three generations of
leptons. 
In general we follow the convention that $\Sl_1$ ($\Sl_2$) has the
large ``left-handed'' (``right-handed'') component.
Besides the symbols equal for all three generations, $\msl1$ and $\msl2$,
we also explicitly use the scalar electron, muon and tau masses,
$\mse{1,2}$, $\msmu{1,2}$ and $\mstau{1,2}$.

Following the stronger experimental limits from the
LHC~\cite{ATLAS-SUSY,CMS-SUSY},
we assume that the colored sector of the MSSM is sufficiently heavier
than the EW sector, and does not play a role in this analysis. For the
Higgs-boson sector we assume that the radiative corrections to the light
$\cp$-even Higgs boson (largely originating from the top/stop sector)
yield a value in agreement with the experimental data, $\Mh \sim 125 \gev$.
This naturally yields stop masses in the TeV
range~\cite{Bagnaschi:2017tru,Slavich:2020zjv}, in agreement 
with the above assumption. Concerning the Higgs-boson mass scale, as
given by the $\cp$-odd Higgs-boson mass, $\MA$, we employ the existing
experimental bounds from the LHC. In the combination with other data,
this results in a non-relevant impact of the heavy Higgs bosons on our
analysis, as was discussed in \citere{CHS1}.


\subsection {Relevant constraints}
\label{sec:constraints}

The experimental result for 
$\amu := \gmin2/2$ was so far dominated by the measurements made at the
Brookhaven National Laboratory (BNL)~\cite{Bennett:2006fi},
resulting in a world average of~\cite{PDG2018}
\begin{align}
\amu^{\rm exp-BNL} &= (11 659 209.1 \pm 6.3) \times 10^{-10}~,
\label{gmt-exp-BNL}
\end{align}
combining statistical and systematic uncertainties.
The SM prediction of \amu\ is given by~\cite{Aoyama:2020ynm}
(based on \citeres{Aoyama:2012wk,Aoyama:2019ryr,Czarnecki:2002nt,Gnendiger:2013pva,Davier:2017zfy,Keshavarzi:2018mgv,Colangelo:2018mtw,Hoferichter:2019mqg,Davier:2019can,Keshavarzi:2019abf,Kurz:2014wya,Melnikov:2003xd,Masjuan:2017tvw,Colangelo:2017fiz,Hoferichter:2018kwz,Gerardin:2019vio,Bijnens:2019ghy,Colangelo:2019uex,Blum:2019ugy,Colangelo:2014qya}
)%
\footnote{
In \citere{CHS1} a slightly different value was used, with a
negligible effect on the results.
}%
, 
\begin{align}
\amu^{\rm SM} &= (11 659 181.0 \pm 4.3) \times 10^{-10}~.
\label{gmt-sm}
\end{align}
Comparing this with the current experimental measurement in \refeq{gmt-exp-BNL}
results in a deviation of
\begin{align}
\Delta\amu^{\rm old} &= (28.1 \pm 7.6) \times 10^{-10}~, 
\label{gmt-diffold}
\end{align}
corresponding to a $3.7\,\sig$ discrepancy.

\smallskip
Recently, the ``MUON G-2'' collaboration~\cite{Grange:2015fou}
at Fermilab published their Run~1 data~\cite{newgmin2}
\begin{align}
\amu^{\rm exp-FNAL} &= (11 659 \fnalcen \pm \fnalunc) \times 10^{-10}~,
\label{gmt-exp-FNAL}
\end{align}
being within $\fnalbnlsig\,\sig$ well compatible with the previous
experimental result in \refeq{gmt-exp-BNL}.
The combined new world average was announced as
\begin{align}
\amu^{\rm exp} &= (11 659 \newcen \pm \newunc) \times 10^{-10}~.
\label{gmt-exp}
\end{align}
Compared with the SM prediction in \refeq{gmt-sm}, one arrives at a new
deviation of
\begin{align}
\Delta\amu &= (\newdiff \pm \newdiffunc) \times 10^{-10}~, 
\label{gmt-diff}
\end{align}
corresponding to a $\newdiffsig\,\sig$ discrepancy.
We use this limit as a cut at the $\pm2\,\sig$ level.
\htrs{Here one should note that the new lower $2\,\sig$ limit is similar
to the old one, which leads to the expectation that the previously upper
bounds, see \citere{CHS1}, are confirmed.}

\medskip
Recently a new lattice calculation for the leading order hadronic
vacuuum polarization (LO HVP) contribution to
$\amu^{\rm SM}$~\cite{Borsanyi:2020mff} has been reported, which,
however, was not used in the new theory world
average, \refeq{gmt-sm}~\cite{Aoyama:2020ynm}. Consequently, we also do
not take this result into account, see also the discussions in
\citeres{CHS1,Lehner:2020crt,Borsanyi:2020mff,Crivellin:2020zul,Keshavarzi:2020bfy,1802677}. 
On the other hand, we are also
aware that our conclusions would change substantially if the result
presented in \cite{Borsanyi:2020mff} turned out to be correct.

\medskip
In MSSM the main contribution to \gmin2\ at the one-loop level comes from
diagrams involving $\cha1-\Sn$ and $\neu1-\tilde \mu$ loops. 
In our analysis the MSSM contribution to \gmin2\
at two loop order is calculated using {\tt GM2Calc}~\cite{Athron:2015rva},
implementing two-loop corrections
from \cite{vonWeitershausen:2010zr,Fargnoli:2013zia,Bach:2015doa}
(see also \cite{Heinemeyer:2003dq,Heinemeyer:2004yq}).

\bigskip
All other constraints are taken into account exactly as
in \citere{CHS1}. These comprise

\begin{itemize}

\item Vacuum stability constraints:\\
All points are checked to possess a stable and correct EW vacuum, e.g.\
avoiding charge and color breaking minima. This check is performed with
the public code {\tt Evade}~\cite{Hollik:2018wrr,Robens:2019kga}.

\item Constraints from the LHC:\\
All relevant EW SUSY searches are taken into account, mostly via
\CM~\cite{Drees:2013wra,Kim:2015wza, Dercks:2016npn}, where many
analysis had to be implemented newly~\cite{CHS1}.

\item
Dark matter relic density constraints:\\
We use the latest result from Planck~\cite{Planck}. The relic density in
the MSSM is evaluated with
\MO~\cite{Belanger:2001fz,Belanger:2006is,Belanger:2007zz,Belanger:2013oya}. 

\item
Direct detection constraints of Dark matter:\\
We employ the constraint on the spin-independent
DM scattering cross-section $\ssi$ from
XENON1T~\cite{XENON} experiment, 
evaluating the theoretical prediction for $\ssi$ using
\MO~\cite{Belanger:2001fz,Belanger:2006is,Belanger:2007zz,Belanger:2013oya}.
A combination with other DD experiments would yield only very slightly
stronger limits, with a negligible impact on our results.

\end{itemize}


\section{Parameter scan}
\label{sec:scan}

We scan the relevant MSSM parameter space to obtain lower and {\it upper}
limits on the relevant neutralino, chargino and slepton masses.
As detailed in \citere{CHS1} three scan regions cover the relevant
parameter space:

\begin{description}
\item
{\bf (A) bino/wino DM with \boldmath{$\chapm1$}-coannihilation}\\
\begin{align}
  100 \gev \leq M_1 \leq 1 \tev \;,
  \quad M_1 \leq M_2 \leq 1.1 M_1\;, \notag \\
  \quad 1.1 M_1 \leq \mu \leq 10 M_1, \;
  \quad 5 \leq \tb \leq 60, \; \notag\\
  \quad 100 \gev \leq \mL \leq 1 \tev, \; 
  \quad \mR = \mL~.
\label{cha-coann}
\end{align}

\item
{{\bf (B) bino DM with \boldmath{$\Slpm$}-coannihilation}}\\
{\bf (B1)} Case-L: SU(2) doublet
\begin{align}
  100 \gev \leq M_1 \leq 1 \tev \;,
  \quad M_1 \leq M_2 \leq 10 M_1 \;, \notag\\
  \quad 1.1 M_1 \leq \mu \leq 10 M_1, \;
  \quad 5 \leq \tb \leq 60, \; \notag\\
  \quad M_1 \gev \leq \mL \leq 1.2 M_1, 
  \quad M_1 \leq \mR \leq 10 M_1~. 
\label{slep-coann-doublet}
\end{align}

{\bf (B2)} Case-R: SU(2) singlet
\begin{align}
  100 \gev \leq M_1 \leq 1 \tev \;,
  \quad M_1 \leq M_2 \leq 10 M_1 \;, \notag \\
  \quad 1.1 M_1 \leq \mu \leq 10 M_1, \;
  \quad 5 \leq \tb \leq 60, \; \notag\\
  \quad M_1 \gev \leq \mR \leq 1.2 M_1,\; 
  \quad M_1 \leq \mL \leq 10 M_1~.
\label{slep-coann-singlet}
\end{align}
\end{description}
In all three scans we choose flat priors of the parameter space and
generate \order{10^7} points.

$\MA$ has also been set to be above the TeV scale. Consequently, we
do not include explicitly the possibility of $A$-pole annihilation,
with $\MA \sim 2 \mneu1$. As we will briefly discuss below the combination of
direct heavy Higgs-boson searches with the other experimental
requirements constrain this possibility substantially.
Similarly, we do not consider $h$-~or $Z$-pole annihilation (see,
e.g., \citere{Carena:2018nlf}), as such a 
light neutralino sector likely overshoots the \gmin2\ contribution, see
the discussion in \citere{CHS1}.


\subsection*{Analysis flow}

The data sample is generated by scanning randomly over the input parameter
range mentioned above, using a flat prior for all parameters.
We use {\tt SuSpect}~\cite{Djouadi:2002ze}
as spectrum and SLHA file generator. The points are required
to satisfy the $\chapm1$ mass limit from LEP~\cite{lepsusy}. 
The SLHA output files
from {\tt SuSpect} are then passed as input to {\tt GM2Calc} and \MO~for
the calculation of \gmin2 and the DM observables, respectively. The parameter
points that satisfy the new \gmin2\ constraint, \refeq{gmt-diff}, the DM
relic density and the direct detection constraints and
the vacuum stability constraints as checked with {\tt Evade} are
then taken to the final 
step to be checked against the latest LHC constraints implemented in \CM,
as described in detail in \citere{CHS1}. The branching ratios of the
relevant SUSY particles are computed using
{\tt SDECAY}~\cite{Muhlleitner:2003vg} 
and given as input to \CM.



\section{Results}
\label{sec:results}

We present the results of the allowed parameter ranges in the three
scenarios defined above. 
We follow the analysis flow as described above
and denote the points surviving certain constraints
with different colors:
\begin{itemize}
\item grey (round): all scan points.
\item green (round): all points that are in agreement with \gmin2, taking
into account the new limit as given in \refeq{gmt-diff}.
\item blue (triangle): points that additionally give the correct relic density,
see \refse{sec:constraints}.
\item cyan (diamond): points that additionally pass the DD constraints,
see \refse{sec:constraints}.
\item red (star):  points that additionally pass the LHC constraints, see \refse{sec:constraints}.
\end{itemize}


\subsection{Results in the three DM scenarios}
\label{sec:res3}

\begin{figure}[htb!]
\begin{center}
\includegraphics[width=0.45\textwidth]{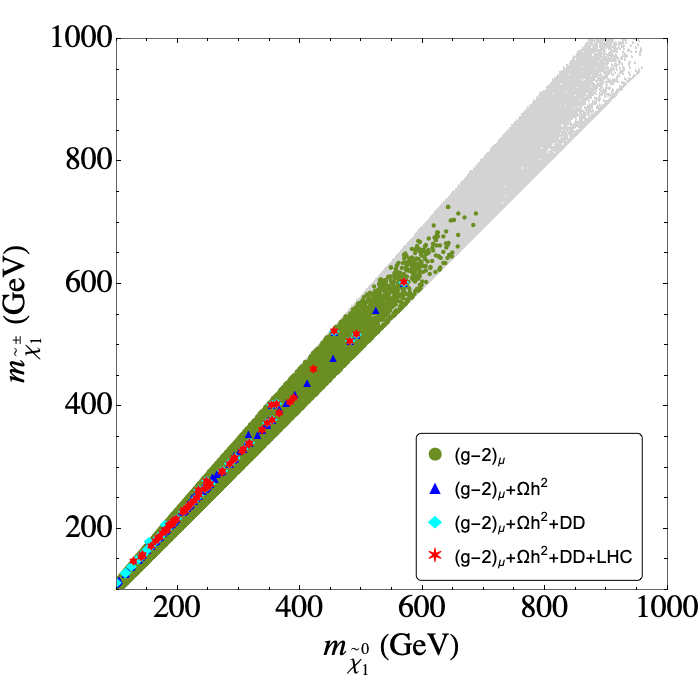}
\includegraphics[width=0.45\textwidth]{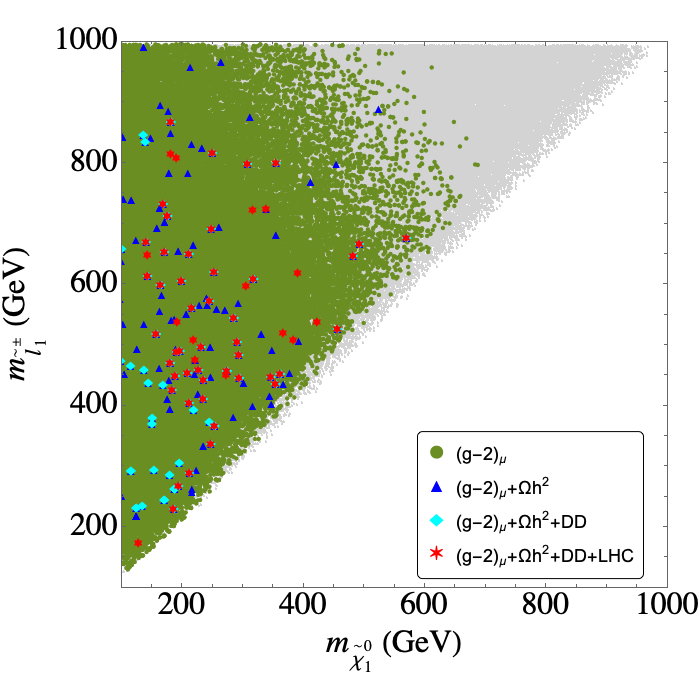}\\
\includegraphics[width=0.45\textwidth]{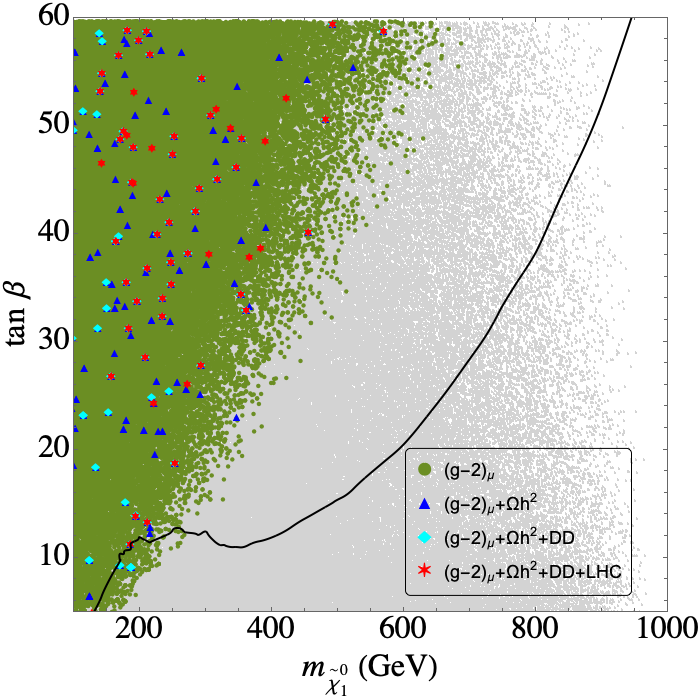}
\caption{\label{fig:charco}
Results in the $\cha1$-coannihilation scenario:
$\mneu1$--$\mcha1$ plane (upper left),
$\mneu1$--$\msl1$ plane (upper right),
$\mneu1$--$\tb$ plane (lower plot);
for the color coding: see text.
}
\end{center}
\end{figure}

We start in \reffi{fig:charco} with the results in the
$\cha1$-coannihilation scenario. 
In the $\mneu1$--$\mcha1$ plane, shown in the upper left plot, by definition
of $\cha1$-coannihilation the points are clustered in the diagonal of
the plane. One observes a clear upper limit on the (green) points
allowed by the new \gmin2\ result of about $\htrs{700} \gev$, which is
similar to the previously obtained one in \citere{CHS1}. 
Applying the CDM constraints reduce the upper limit further to
about $\htrs{600} \gev$, again similar as for the old \gmin2\ result.
Applying the LHC
constraints, corresponding to the ``surviving'' red points (stars), does
not yield a further reduction from above, but cuts always 
only points in the lower mass range.
Thus, the new experimental data set an upper as well as a lower bound,
yielding a clear search target for the upcoming LHC runs, and in
particular for future $e^+e^-$ colliders, as will be briefly discussed
in \refse{sec:future}. 

The distribution of the lighter slepton mass (where it should be kept in
mind that we have chosen the same masses for all three generations,
see \refse{sec:model}) is presented in the $\mneu1$-$\msl1$ plane, shown
in the upper right plot of \reffi{fig:charco}. 
The \gmin2\ constraint places important constraints in this mass plane,
since both types of masses enter into the contributing SUSY diagrams,
see \refse{sec:constraints}. 
The constraint is satisfied in a triangular region with its tip
around $(\mneu1, \msl1) \sim (\htrs{700} \gev, \htrs{800} \gev)$, 
compatible with the old limits.
This is slightly reduced to $\sim (600 \gev, 700 \gev)$
when the DM constraints are
taken in to account, as can be seen in the distribution of
the blue, cyan and red points (triangle/diamond/star).
The LHC constraints cut out lower slepton masses, going up to
$\msl1 \lsim 400 \gev$, as well as part of the very low $\mneu1$
points nearly independent of $\msl1$. Details on these cuts can be found
in \citere{CHS1}. 

We finish our analysis of the $\cha1$-coannihilation case with the
$\mneu1$-$\tb$ plane presented in the lower plot of \reffi{fig:charco}.
The \gmin2\ constraint is
fulfilled in a triangular region with largest neutralino masses allowed
for the largest $\tb$ values (where we stopped our scan at $\tb = 60$).
In agreement with the previous plots, the largest values for the
lightest neutralino masses are $\sim \htrs{600} \gev$, as before 
compatible with the old \gmin2\ limit~\cite{CHS1}. 
The LHC constraints cut out points at low $\mneu1$, but nearly
independent on $\tb$.
In this plot we also show as a black line the current bound
from LHC searches for heavy neutral Higgs
bosons~\cite{Bahl:2018zmf} in the channel $pp \to H/A \to \tau\tau$
in the $M_h^{125}(\tilde\chi)$ benchmark scenario
(based on the search data published
in \citere{Aad:2020zxo} using $139\, \ifb$.)%
\footnote{We thank T.~Stefaniak for providing us the limit, using
the latest version of
{\tt HiggsBounds}~\cite{Bechtle:2008jh,Bechtle:2011sb,Bechtle:2013wla,Bechtle:2015pma,Bechtle:2020pkv}.}%
.
~The black line corresponds to
$\mneu1 = \MA/2$, i.e.\ roughly to the requirement for $A$-pole
annihilation, where points
above the black lines are experimentally excluded. 
There are a very few  passing the current \gmin2\ constraint
below the black $A$-pole line, reaching up to $\mneu1 \sim \htrs{250} \gev$,
for which the $A$-pole annihilation could provide the correct DM relic
density. It can be expected that with the improved limits as given
in \cite{Aad:2020zxo} this possibility is further restricted. These
effects makea the $A$-pole annihilation in this scenario marginal.

The final parameter constrained in this scenario is the Higgs-mixing
parameter~$\mu$. Here in particular the DD bounds are
important. Following the analysis in \citere{CHS1} we find a lower
limit of $\mu/M_1 \gsim \htrs{1.8}$ in the $\cha1$-coannihilation
scenario.

\begin{figure}[htb!]
\begin{center}
\includegraphics[width=0.45\textwidth]{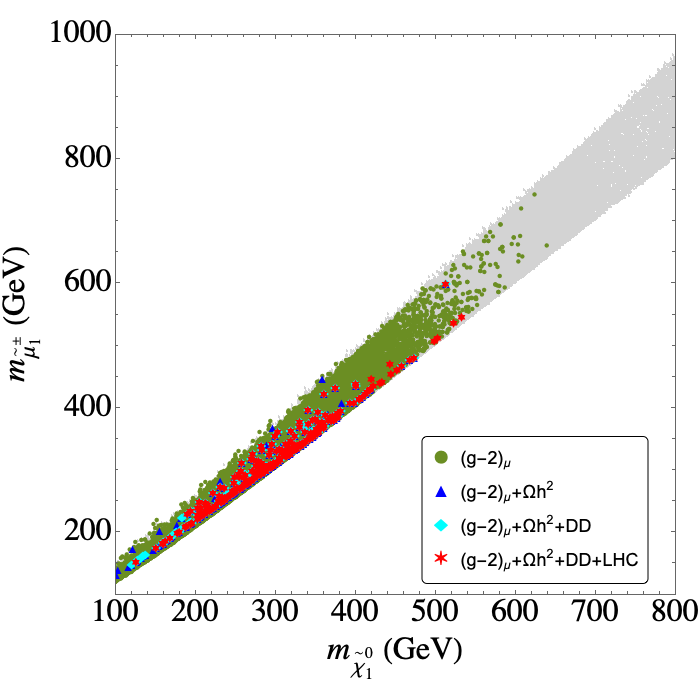}
\includegraphics[width=0.45\textwidth]{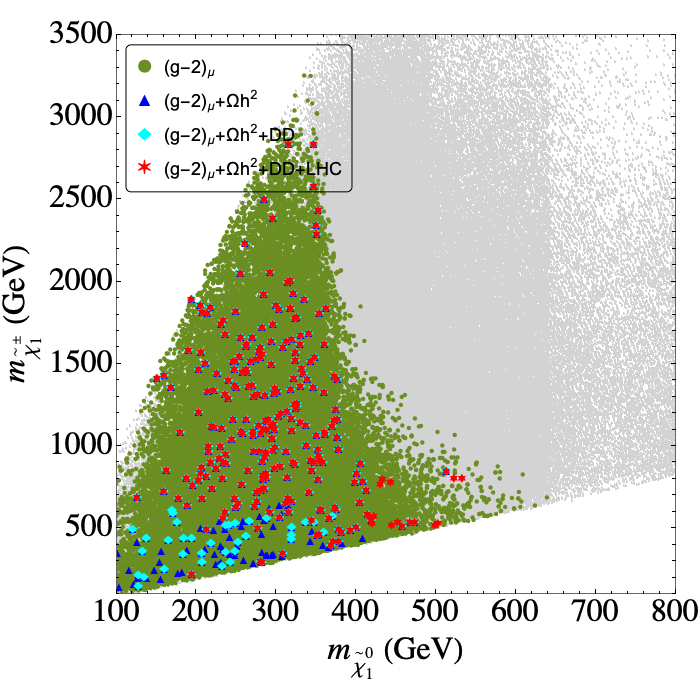}\\
\includegraphics[width=0.45\textwidth]{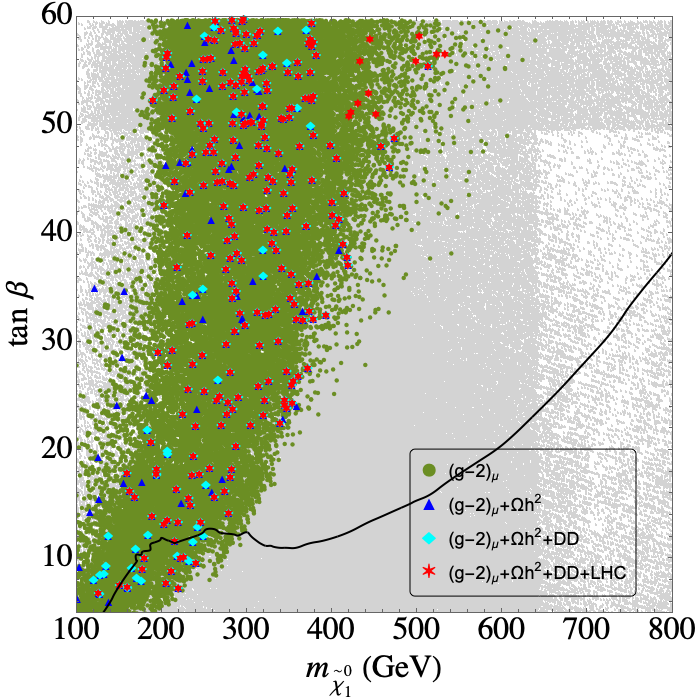}
\caption{\label{fig:slepco-L}
Results in the $\Slpm$-coannihilation Case-L:
$\mneu1$--$\msl1$ plane (upper left),
$\mneu1$--$\mcha1$ plane (upper right),
$\mneu1$--$\tb$ plane (lower plot);
for the color coding: see text.
}
\end{center}
\end{figure}

\medskip
We now turn to the case of $\Slpm$-coannihilation Case-L, as shown
in \reffi{fig:slepco-L}. We start with the $\msmu1 -\mneu1$ plane in the
upper left plot. 
By definition of the scenario, the points are located along the
diagonal of the plane.
The new constraint from \gmin2\ puts an upper bound of
$\sim \htrs{650} \gev$ on the masses, which is in the same ballpark as for
the old \gmin2\ results~\cite{CHS1}. 
Including the DM and LHC
constraints, the bound is reduced to $\sim \htrs{540} \gev$, again compatible
with \citere{CHS1}.
As in the case of $\cha1$-coannihilation the LHC constraints cut away
only low mass points. 
The corresponding implications for the searches at future colliders
are briefly discussed in \refse{sec:future}. 

In upper right plot of \reffi{fig:slepco-L} we show the results in the
$\mneu1$-$\mcha1$ plane. The \gmin2\ limits on $\mneu1$ become
slightly stronger for larger chargino masses, and upper limits on the
chargino mass are set at $\sim \htrs{2.8} \tev$.
The LHC limits cut away a lower wedge going up to
$\mcha1 \lsim 600 \gev$.

The results for the $\Slpm$-coannihilation Case-L in the
$\mneu1$-$\tb$ plane are presented in the lower plot of
\reffi{fig:slepco-L}. The
overall picture is similar to the $\cha1$-coannhiliation case shown
above in \reffi{fig:charco}. Larger LSP masses are allowed for
larger $\tb$ values. On the other hand the combination of small $\mneu1$
and large $\tb$ leads to a {\it too large} contribution to
$\amu^{\rm SUSY}$ and is thus excluded. As in \reffi{fig:charco}
we also show the limits from $H/A$ searches at the LHC, where we set
(as above) $\mneu1 = \MA/2$, i.e.\ roughly to the requirement for $A$-pole
annihilation, where points above the black lines are experimentally excluded. 
In this case for the current \gmin2\ limit substantially more
points passing  the \gmin2\ constraint ``survive'' below the black
line, i.e.\ they are potential candidates for $A$-pole
annihilation. The masses reach up to $\sim \htrs{240} \gev$.
Together with the already
stronger bounds on $H/A \to \tau\tau$~\cite{Aad:2020zxo} this does not
fully exclude $A$-pole annihilation, but leaves it as a rather remote
possibility.

The limits on $\mu/M_1$ (not shown) in the $\Slpm$-coannihilation Case-L
are again mainly driven by the DD-experiments. 
Given both CDM constraints and the LHC
constraints, the smallest $\mu/M_1$ value we find is~\htrs{1.7}.

\begin{figure}[htb!]
\begin{center}
\includegraphics[width=0.45\textwidth]{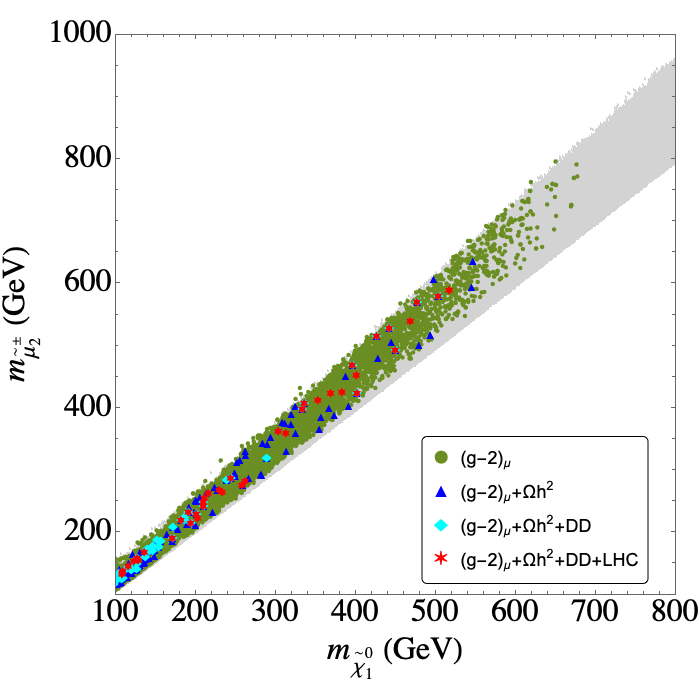}
\includegraphics[width=0.45\textwidth]{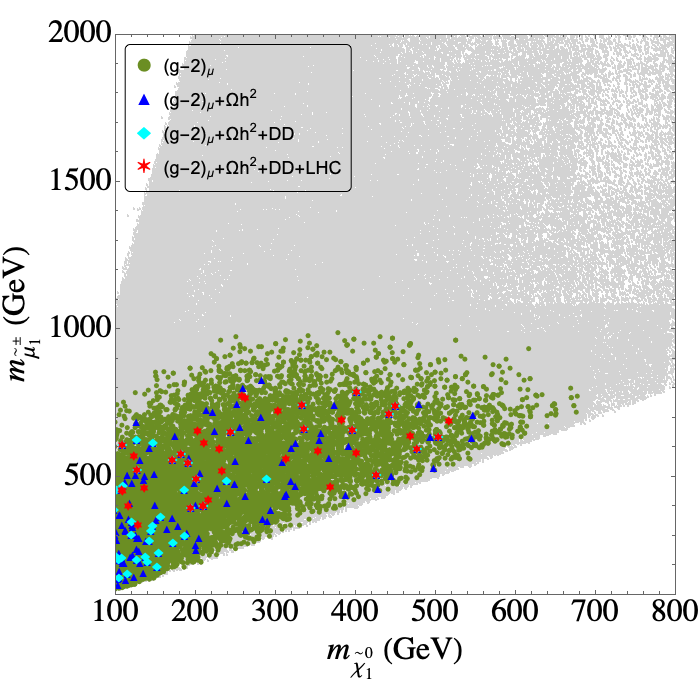}\\
\includegraphics[width=0.45\textwidth]{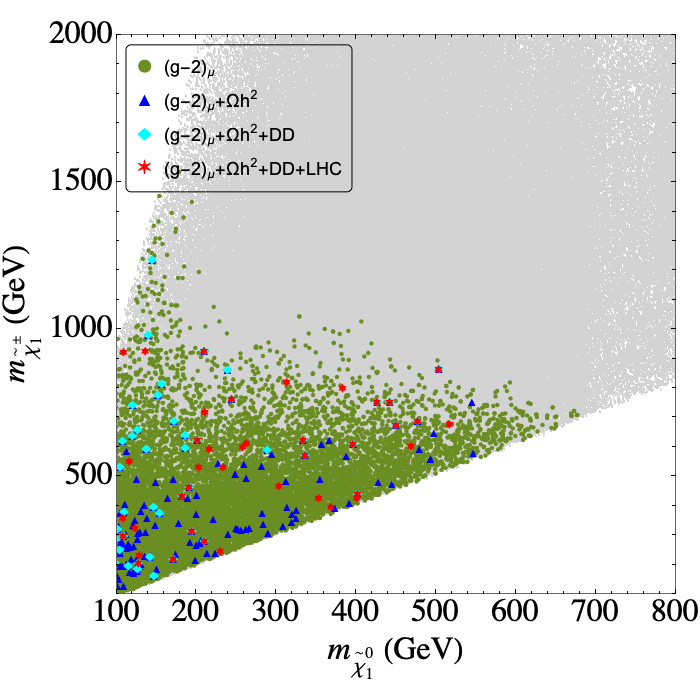}
\includegraphics[width=0.45\textwidth]{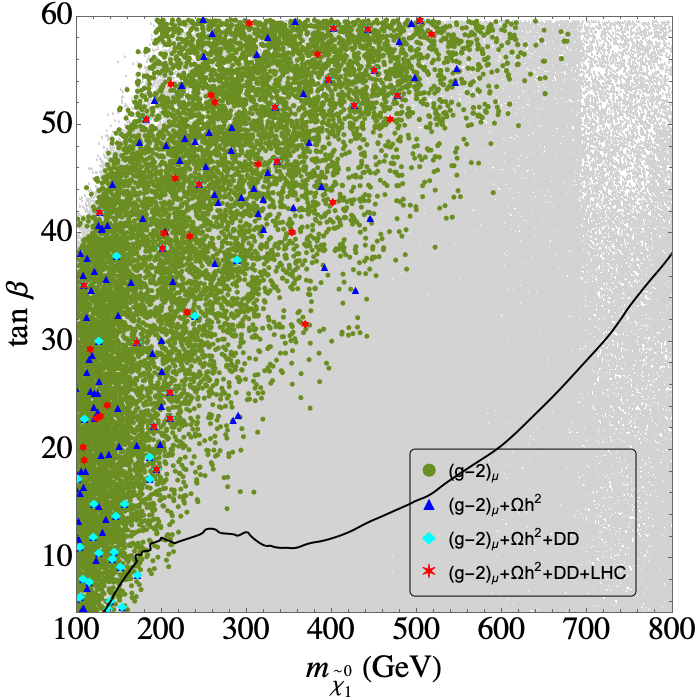}
\caption{\label{fig:slepco-R}
Results in the $\Slpm$-coannihilation Case-R.
$\mneu1$--$\msl2$ plane (upper left),
$\mneu1$--$\msl1$ plane (upper right),
$\mneu1$--$\mcha1$ plane (lower left),
$\mneu1$--$\tb$ plane (lower right plot);
for the color coding: see text.
}
\end{center}
\end{figure}

\medskip
We now turn to our third scenario, $\Slpm$-coannihilation Case-R,
where in the scan we require the ``right-handed'' sleptons to be close
in mass with the LSP. It should be kept in mind that in our notation
we do not mass-order the sleptons: for negligible mixing as it is
given for selectrons and smuons the ``left-handed'' (``right-handed'')
slepton corresponds to $\Sl_1$ ($\Sl_2$).
We start in \reffi{fig:slepco-R} with the $\mneu1$-$\msmu2$ plane
in the upper left plot. By definition
of the scenario the points are concentrated on the diagonal. The new
\gmin2\ bound yields an upper limit on the LSP of
$\sim \htrs{690} \gev$, a small reduction w.r.t.\ the
previous results~\cite{CHS1}. The new \gmin2\ bound also places
an upper limit on $\msmu2$ (which is close
in mass to the $\Sel2$ and $\Stau2$) of $\sim \htrs{800} \gev$, again
in the same ballpark as for the old \gmin2\ result.
Including the CDM and LHC constraints, these limits reduce to
$\sim \htrs{520} \gev$ for the LSP, 
and correspondingly to $\sim \htrs{600} \gev$ for $\msmu2$,
and $\sim \htrs{530} \gev$ for $\mstau2$.
The LHC constraints cut out some, but not all lower-mass
points. 

The distribution of the heavier slepton is displayed in the
$\mneu1$-$\msmu1$ plane in the upper right plot of \reffi{fig:slepco-R}.
Although the ``left-handed'' sleptons are allowed to be much heavier,
the \gmin2\
constraint imposes an upper limit of $\sim \htrs{950} \gev$, about
$\sim \htrs{50} \gev$ less than previously. This effect is discussed in
detail in \citere{CHS1}. The DM and LHC constraints to not yield a
further reduction in this case, which cut away only lower mass points
and set a lower limit of $\sim 300 \gev$ for the heavier sleptons in the
Case-R.

In the lower left plot of \reffi{fig:slepco-R} we show the results in the
$\mneu1$-$\mcha1$ plane. As in the Case-L the \gmin2\ limits on
$\mneu1$ become slightly stronger for larger chargino masses.
The upper limits on the
chargino mass, however, are substantially stronger as in the Case-L. They are
reached at $\sim \htrs{900} \gev$ using the new \gmin2\ result, similar to
the limits for the old \gmin2\
limit~\cite{CHS1}. 

We finish our analysis of the $\Slpm$-coannihilation Case-R  with the
results in the $\mneu1$-$\tb$ plane, presented in the lower right plot
of \reffi{fig:slepco-R}. The
overall picture is similar to the previous cases shown above.
Larger LSP masses are allowed for larger $\tb$ values. On the other
hand the combination of small $\mneu1$ and very large $\tb$ values,
$\tb \gsim 40$ leads to stau masses below the LSP mass, which we
exclude for the CDM constraints.
The LHC searches mainly affect parameter points with $\tb \lsim 20$. 
Larger $\tb$ values induce a larger mixing in the third
slepton generation, enhancing the probability for charginos to
decay via staus and thus evading the LHC constraints.
As above we also show the limits from $H/A$ searches at the LHC, where we set
(as above) $\mneu1 = \MA/2$, i.e.\ roughly to the requirement for $A$-pole
annihilation, where points above the black lines are experimentally excluded. 
Comparing Case-R and Case-L 
substantially less points are passing the new \gmin2\ constraint
below the black line, i.e.\ are potential candidates for $A$-pole
annihilation. 
The masses reach only up to $\sim \htrs{150} \gev$, about
$\sim \htrs{50} \gev$ less than with the old \gmin2\ result.
Together with the already
stronger bounds on $H/A \to \tau\tau$~\cite{Aad:2020zxo} this leaves 
$A$-pole annihilation as a quite remote
possibility in this scenario.

The limits on $\mu/M_1$ (not shown) in the $\Slpm$-coannihilation Case-R
are as before mainly driven by the DD-experiments. 
Given both CDM constraints and the LHC
constraints, the smallest $\mu/M_1$ value we find is~\htrs{1.7}.


\subsection{Implications for future colliders}
\label{sec:future}

In \citere{CHS1} we had evaluated the prospects for EW searches at the
HL-LHC~\cite{CidVidal:2018eel} and at a
hypothetical future $e^+e^-$ collider such as
ILC~\cite{ILC-TDR,LCreport} or CLIC~\cite{CLIC,LCreport}.  

The prospects for BSM phenomenology at the HL-LHC have been summarized
in \cite{CidVidal:2018eel} for a 14 TeV run with 3 \iab\ of
integrated luminosity. 
For the direct production of charginos and neutralinos through EW
interaction, the projected  95\% exclusion reach as well as a 5$\sigma$
discovery reach have been presented. Following the discussion
in \citere{CHS1} we conclude that via these searches
the updated \gmin2\ limit together with DM constraints can conclusively
probe ``almost'' the entire allowed 
parameter region of $\Slpm$-coannihilation Case-R scenario
and a significant part of the same parameter space
for Case-L scenario at the HL-LHC. 
On the other hand, the analysis
for compressed higgsino-like spectra at the HL-LHC, see the discussion
in \citere{CHS2}, may exclude
$\mneu2 \sim \mcha1 \sim 350 \gev$ with mass gap as low as 2~GeV for
$\mcha1$ around 100~GeV. Hence, a substantial parameter region can be
curbed for the $\cha1$-coannihilation case in the absence of a signal in
the compressed scenario analysis with soft leptons at the final state.
However, higher energies in $pp$ collisions or an $e^+e^-$ collider
with energies up to $\sqrt{s} \sim 1.5 \tev$ will be needed to probe
this scenario completely~\cite{Strategy:2019vxc,Berggren:2020tle}).

\medskip
Direct production of EW particles at $e^+e^-$ colliders clearly
profits from a higher center-of-mass energy, $\sqrt{s}$. Consequently,
we focus here on the two proposals for linear $e^+e^-$ colliders,
ILC~\cite{ILC-TDR,LCreport} and CLIC~\cite{CLIC,LCreport}, which can
reach energies up to $1 \tev$, and $3 \tev$, respectively.
In \citere{CHS1} we had evaluated the cross-sections for the various
SUSY pair production modes (based on
~\citeres{Heinemeyer:2017izw,Heinemeyer:2018szj}) for the energies
currently foreseen in the run plans of the two colliders.

Taking into account the results for the cross-sections evaluated
in \citere{CHS1}, one can conclude that the new accuracy on \gmin2,
yielding similar upper limits on EW SUSY 
particles, guarantees the discovery 
at the higher-energy stages of the ILC and/or CLIC.
This holds in particular for the LSP and the NLSP.
The improved \gmin2\ constraint, confirming the deviation of $\amu^{\rm exp}$
from the SM prediction, clearly strengthens the case for future $e^+e^-$
colliders.

\smallskip
As discussed in \refse{sec:scan} we have not considered the
possibility of $Z$~or $h$~pole annihilation to find agreement of the
relic DM density with the other experimental measurements. 
However, it should be noted that in this context an LSP with
$M_1 \sim \mneu1 \sim \MZ/2$ or $\sim \Mh/2$ would yield a detectable
cross-section $e^+e^- \to \neu1\neu1\ga$ in any future high-energy $e^+e^-$
collider. Furthermore, depending on the values of $M_2$ and $\mu$,
this scenario likely yields other clearly detectable EW-SUSY
production cross-sections at future $e^+e^-$ colliders. We leave this
possibility for future studies.

On the other hand, the possibility of $A$-pole annihilation was briefly
discussed for all three scenarios. While it appears a rather remote
possibility, it cannot be fully excluded by our analysis. However,
even in the ``worst'' case of $\Slpm$-coannihilation Case-L an upper
limit on $\mneu1$ of $\sim \htrs{250} \gev$ can be set. While not as low as
in the case of $Z$~or $h$-pole annihilation, this would still offer
good prospects for future $e^+e^-$ colliders. We leave also this
possibility for future studies.


\section {Conclusions}
\label{sec:conclusion}

The electroweak (EW) sector of the MSSM, consisting of charginos,
neutralinos and scalar leptons can account for a variety of experimental
data: the CDM relic abundance with the lightest neutralino, $\neu1$ as
LSP, the bounds from
DD experiments as well as from direct searches at the LHC. Most
importantly, the EW sector of the 
MSSM can account for the long-standing discrepancy of \gmin2.
The new result for the Run~1 data of the ``MUON G-2'' experiment
confirmed the deviation from the SM prediction found previously. 

Under the assumption that the previous experimental result on \gmin2\ is
uncorrelated with the new ``MUON G-2'' result we combined the data and
obtained a new deviation from the SM prediction of
$\Delta\amu = (\newdiff \pm \newdiffunc) \times 10^{-10}$, 
corresponding to a $\newdiffsig\,\sig$ discrepancy.
We used this limit as a cut at the $\pm2\,\sig$ level.

In this paper, under the assumption that the $\neu1$
provides the full DM relic abundance we analyzed which mass ranges of
neutralinos, charginos and sleptons are in agreement with all
relevant experimental data: the new limit for \gmin2\,, the relic
density bounds, the DD experimental bounds, as well as the LHC searches
for EW SUSY particles. These results present an update
of \citere{CHS1}, where the previous \gmin2\ result had been used (as
well as a hypothetical ``MUON G-2'' result). 

We analyzed three scenarios, depending on the mechanism that brings the
relic density in agreement with the experimental data:
$\cha1$-coannihilation, $\Slpm$-coannihilation with the mass of the
``left-handed'' (``right-handed'') slepton close to $\mneu1$, Case-L
(Case-R). We find in all three cases a clear upper limit on $\mneu1$.
We find that the upper limits on the LSP mass are decreased
to about $\htrs{600} \gev$ for 
$\chapm1$-coannihilation, $\htrs{540} \gev$ for $\Slpm$-coannihilation Case-L
and $\htrs{520} \gev$ in Case-R,
confirming the collider targets w.r.t.\ the old \gmin2.
Similarly, the upper limits on the NLSP masses are confirmed to about
$\htrs{600} \gev$, $\htrs{600} \gev$ and $\htrs{530} \gev$ in the three
cases that we have 
explored, again compatible with the
previous \gmin2\ result.

For the HL-LHC we have briefly discussed the prospects to cover the
parameter regions that are preferred by the new \gmin2\ result.
In particular the $\Slpm$-coannihilation Case-R can be conclusively
tested at the HL-LHC, while the other scenarios are only partially
covered. 
Concerning future high(er) energy $e^+e^-$ colliders, ILC and CLIC, 
one can conclude that the new accuracy on \gmin2\
confirms the upper limits on EW SUSY, and it can be expected
that at least some particles can 
be discovered at the higher-energy stages of the ILC and/or CLIC.
This holds in particular for the LSP and the NLSP. 
Therefore, the new \gmin2\ constraint, confirming the deviation
of $\amu^{\rm exp}$ from the SM prediction, strongly motivates
the need of future $e^+e^-$ colliders.


\subsection*{Acknowledgments}

I.S.\ gratefully thanks S.~Matsumoto for the cluster facility.
The work of I.S.\ is supported by World Premier
International Research Center Initiative (WPI), MEXT, Japan.
The work of S.H.\ is supported in part by the
MEINCOP Spain under contract PID2019-110058GB-C21 and in part by
the AEI through the grant IFT Centro de Excelencia Severo Ochoa SEV-2016-0597.
The work of M.C.\ is supported by the project AstroCeNT:
Particle Astrophysics Science and Technology Centre,  carried out within
the International Research Agendas programme of
the Foundation for Polish Science financed by the
European Union under the European Regional Development Fund.



\newcommand\jnl[1]{\textit{\frenchspacing #1}}
\newcommand\vol[1]{\textbf{#1}}

\newpage{\pagestyle{empty}\cleardoublepage}


\end{document}